\documentclass[runningheads]{llncs}
\usepackage{amsmath}
\usepackage{amssymb}
\usepackage{placeins}
\usepackage{bbding}
\usepackage[T1]{fontenc}
\usepackage{graphicx}
\usepackage{hyperref}
\usepackage{color}

\newcommand{\ie}{i.e., }
\usepackage{multirow}
\begin{document}
\title{An Accessible Solution for Deformable Image Registration Compared with Learning-Based Approaches}
\titlerunning{Accessible DIR vs. Learning-Based Approaches}
\author{Onur Ali Zeybekoglu\inst{1,2}\, \and
David Tilly\inst{2,3} \and
Orcun Goksel\inst{1}}

\authorrunning{Zeybekoglu et al.}

\institute{
Department of Information Technology, Uppsala University, Uppsala, Sweden\\
\and
Department of Medical Physics, Uppsala University Hospital, Uppsala, Sweden
\and
Department of Immunology, Genetics and Pathology, Uppsala University, Uppsala, Sweden
}

\maketitle              
\begin{abstract}
Deformable image registration (DIR) is a core problem in medical image analysis; but, unlike labeling decision problems such as classification and segmentation, registration is a problem class that involves stringent physical constraints. 
Although deep learning methods have made faster registration possible, the resulting models are often difficult to interpret compared to hand-crafted methods with explicit objectives and interpretable physical meaning.
In this work, we show that an analytical method can still yield competitive and superior results to deep learning in a common deformable registration task.
We study pTVreg as a parametric total variation based registration in that context.
Observing its different implementations to perform at various degrees, we introduce here an accessible implementation of this method, together with a Bayesian optimization framework that automatically sets self-parameters for any DIR task from a set of sample examples. 
Experiments on Lung250M-4B show that our proposed implementation achieves state-of-the-art results in this benchmark, substantially superior to existing deep learning solutions and other pTVreg variants as baselines. The source code will be made publicly available at \url{https://github.com/oazeybekoglu/ptvreg-python}.
\end{abstract}
\section{Introduction}
\label{sec:intro}
Image registration plays a central role in many medical image analysis tasks, such as radiation therapy and treatment planning, image-guided surgery, longitudinal disease monitoring, atlas-based segmentation, and multimodal image fusion. Its goal is to establish spatial correspondence between a fixed image and a moving image by estimating a transformation that optimizes a similarity measure between them. 

In deformable image registration, the underlying transformation is typically modeled as a continuous flow field, which can be estimated using either classical optimization-based approaches or more recent deep learning-based methods. CNN-based registration methods, such as VoxelMorph~\cite{balakrishnan2019voxelmorph}, and other subsequent learning-based approaches~\cite{dalca2019unsupervised,mok2020lapirn}, have demonstrated strong performance, particularly in terms of inference speed compared to conventional iterative methods. More recently, transformer-based registration models have been introduced to address the limited receptive field of CNNs and to better capture long-range spatial dependencies, with some examples including TransMorph~\cite{chen2022transmorph} and H-ViT~\cite{ghahremani2024hvit}. However, applying deep learning to deformable image registration, especially for inhale--exhale lung CT, remains challenging because DIR is a highly non-convex dense regression problem. In this context, classical variational and optimization-based methods~\cite{vishnevskiy2017isotropic,rueckert1999nonrigid,ruhaak2017estimation} remain attractive due to their interpretability, explicit control over regularization, and consistently strong performance. At the same time, their use can be limited by implementation complexity and by the need to adapt a small number of important hyperparameters to each new dataset. 

One of the state-of-the-art methods has been based on the isotropic total-variation regularization with an iterative coarse-to-fine solution of linear B-spline deformation grids~\cite{vishnevskiy2017isotropic}, herein referred to as \emph{pTVreg}.
Several hyperparameters, such as the regularization weight, control-point spacing, and the local normalized cross-correlation (image metric) kernel size, were identified in~\cite{vishnevskiy2017isotropic} as having a major effect on the registration results, while reporting them to be relatively stable across different datasets of \emph{intra-patient breathing-motion in CT}. Nevertheless, it is not clear whether these same settings would work reliably for any registration setting, including substantially different tasks such as inter-patient registration of a different modality. 

In this work, we analyze and improve pTVreg, showing state-of-the-art results with a new implementation in Python by incorporating an alternating direction method of multipliers (ADMM) based optimization scheme, testing higher-order interpolations for warping, and also facilitating an easier adaptation of this method to new tasks/datasets by introducing an automatic Bayesian-optimization based parameter-tuning in the framework itself. 
We evaluate our proposed solution comparatively on the Lung250M-4B benchmark for intra-patient lung CT registration~\cite{falta2023lung250m}.
\section{Method}
\label{sec:method}

\subsection{Analytical Formulation of Deformable Image Registration}
\label{sec:problem_formulation}

Let $\Omega$ denote the discrete image domain and let $I_f : \Omega \rightarrow \mathbb{R}$ and $I_m : \Omega \rightarrow \mathbb{R}$ be the fixed and moving images, respectively. Deformable image registration seeks spatially-varying transformation $\phi : \Omega \rightarrow \Omega$ that aligns the moving image with the fixed image. Throughout this work, we use the fixed-domain convention, \ie for each location $x \in \Omega$, $\phi(x)$ denotes the corresponding location in the moving image. The warped moving image is therefore written as
\[
I_m^\phi(x) := (I_m \circ \phi)(x) = I_m(\phi(x)).
\]
It is convenient to express transformations in terms of a flow field $u : \Omega \rightarrow \mathbb{R}^d$, i.e. $\phi(x) = x + u(x)$, where the warped moving image then becomes
\[
I_m^\phi(x) = I_m(x + u(x)).
\]
A standard formulation of deformable registration is hence given by
\begin{equation}
\hat{u}
=
\arg\min_{u}
\mathcal{D}\bigl(I_f, I_m(x+u(x))\bigr)
+
\lambda \mathcal{R}(u),
\label{eq:dense_dir_objective}
\end{equation}
where $\mathcal{D}$ denotes an image similarity term, $\mathcal{R}$ is a regularization term, and $\lambda > 0$ controls the trade-off between the two.  

This shows the two important requirements of registration: the warped moving image should match the fixed image, while the estimated deformation should remain anatomically plausible. Since many different deformations can result in similar image similarity values, registration is an ill-posed problem and hence requires regularization for a robust solution.

\subsection{Image Metric}
\label{sec:Imagemetric}

The data term $\mathcal{D}$ measures the agreement between the fixed image $I_f$ and the warped moving image $I_m^\phi$.
For mono-modal registration, a standard choice is sum of squared differences (SSD), which in the discrete setting is given by
\begin{equation}
\mathcal{D}_{\mathrm{SSD}}(I_f,I_m^\phi) =
 \sum_{x \in \Omega}
\left(I_f[x]-I_m^\phi[x]\right)^2,
\end{equation}

To better handle local intensity variations, an alternative robust metric is the local normalized cross-correlation (LCC)~\cite{cachier2000nonrigid,cachier2003iconic} within Gaussian-windowed patches. 
Let $H_w$ denote a spatially invariant Gaussian weighting kernel with standard deviation $w$, defining the local weighted averages and covariance as:
\begin{equation}
\bar{I} = H_w \ast I,
\qquad
\sigma^2(I) = \overline{I^2} - \bar{I}^{\,2}, \qquad
\langle I_1, I_2 \rangle = \overline{I_1 \odot I_2} - \bar{I}_1 \odot \bar{I}_2,
\end{equation}
where $\ast$ denotes convolution and $\odot$ denotes element-wise multiplication. The corresponding LCC-based dissimilarity can then be written as
\begin{equation}
\mathcal{D}_{\mathrm{LCC}}(I_f,I_m^\phi)
= -\sum_{x \in \Omega}
\frac{\langle I_f, I_m^\phi \rangle[x]}{
\sqrt{\sigma^2(I_f\phantom{^\phi}\!\!\!)[x]+\varepsilon}\ 
\sqrt{\sigma^2(I_m^\phi)[x]+\varepsilon}
},
\end{equation}
where a small constant $\varepsilon>0$ is added inside the square roots for numerical stability in low-contrast regions. This formulation avoids explicit patch-wise computation and can be implemented very efficiently using convolutions. The kernel bandwidth $w$ determines the effective locality of the similarity measure and is therefore an important metric parameter.

\subsection{Regularization}

The regularization term $\mathcal{R}$ imposes prior assumptions on the flow field $u$. A widely used regularization penalizes the squared $\ell_2$ norm of the spatial derivatives of displacement components~\cite{rueckert1999nonrigid}, \ie
$\mathcal{R}_{\mathrm{L2}}(u) = \sum_{j=1}^{d} \|\nabla u_j\|_2^2.$
Despite being smooth and easy to optimize, this may oversmooth motion discontinuities. 
This can be addressed by using total variation (TV)~\cite{rudin1992nonlinear}, which in an anisotropic form penalizes displacement gradient magnitudes component-wise independently
$
\mathcal{R}_{\mathrm{aTV}}(u)=\sum_{j=1}^{d} \|\nabla u_j\|_1\,,
$
which however often causes axis-aligned artifacts.
In this work we use an isotropic TV formulation~\cite{blomgren1998colortv}, which couples all spatial derivatives of all displacement components.
Since this becomes non-differentiable at zero, a typical approximation is to make it quadratic locally around zero with a margin of $\epsilon$ as follows:
\begin{equation}
\mathcal{R}_{\mathrm{iTV}}(u)
= \|D(u)\|_{2,1}
\approx \sum_{x \in \Omega}
\sqrt{\sum_{i=1}^{d}\sum_{j=1}^{d} \left(\nabla_i u_j[x]\right)^2 + \epsilon}
\end{equation}
with
$D(u)=\begin{bmatrix} \nabla_1 u_1 & \nabla_2 u_1 & \cdots & \nabla_d u_d \end{bmatrix}^{\!\top}$.
Unlike quadratic penalties, isotropic TV enforces piecewise-smooth flow fields while still allowing sharp local transitions. This is particularly useful in the presence of sliding anatomical motion, but is also advantageous for other non-smooth settings such as inter-patient registration. 

\subsection{Parametrizing the Flow Field into a Deformation Grid}
\label{sec:parametric_model}
Since optimizing a dense flow field can be computationally intensive, often the deformation is parameterized~\cite{rueckert1999nonrigid,schwarz2007nonrigid} on a coarse set of control-point displacements $k$ defined on a Cartesian deformation grid with spacing $K_i$ along axis $i$ in voxels. In the isotropic case, this reduces to a single spacing parameter $K$, with $K_i=K$ for all axes.
Flow field can then be obtained using an interpolation operator $\mathcal{B}$, for which we use a linear (first-order B-spline) approach due to its theoretical guarantees by giving bounds on the flow field~\cite{vishnevskiy2017isotropic}:
\[
u(x) = \mathcal{B}(k)(x),
\qquad
\phi_k(x) = x + \mathcal{B}(k)(x).
\]
With this parametrization, the regularization is imposed on the control-point displacements rather than directly on the flow field. 

Then, the optimization of Eq.~\eqref{eq:dense_dir_objective} to find the deformations can be written using the parametric grid for isotropic TV-regularization:
\begin{equation}
\hat{k} = \arg\min_k
\mathcal{D}\bigl(I_f, I_m \circ \phi_k\bigr) + \lambda \eta \|D(k)\|_{2,1},
\label{eq:parametric_ptv_objective}
\end{equation}
where $\eta$ denotes the control-grid cell volume, \ie
$
\eta = \prod_{i=1}^{d} \delta_i K_i,
$
with voxel spacing $\delta_i$.

The above formulation reduces the solution-space degrees-of-freedom and introduces an additional form of implicit regularization, as the resulting dense flow field is constrained by the chosen control grid and interpolation model. 
Such parametric models can be effectively combined with a coarse-to-fine strategy, where registration is first performed at a coarse scale to capture large motions and then progressively refined at finer scales to recover local details.

\subsection{Optimization}
\label{sec:admm}
The above can be solved using ADMM much more efficiently compared to conventional gradient-based approaches.

In the parametric setting, the regularization is imposed on the control-point displacements $k$ rather than directly on the dense flow field $u$. Accordingly, the parametric isotropic TV term is written as $\eta\|D(k)\|_{2,1}$, where $D$ denotes the discrete spatial derivative operator on the control grid and $\eta$ is the volume of one control-grid cell.

When TV regularization is used, the resulting objective~\eqref{eq:parametric_ptv_objective} combines a smooth but generally non-convex data term with a convex but non-differentiable regularization term. This makes direct gradient-based optimization difficult. To address this issue, the alternating direction method of multipliers (ADMM)~\cite{boyd2011admm,figueiredo2012algorithms} is proposed in pTVreg~\cite{vishnevskiy2017isotropic} but not implemented in the released code, which separates the smooth data term from the non-smooth regularization term.

To decouple the non-differentiable term from the data term, we introduce an auxiliary variable $Z$ such that
$
Z = D(k).
$
This yields the constrained problem
\begin{equation}
\min_{k,Z}
\mathcal{D}\bigl(I_f, I_m \circ \phi_k\bigr)
+
\lambda \eta \|Z\|_{2,1}
\qquad
\text{subject to}
\qquad
D(k) = Z.
\end{equation}

Using the scaled form of ADMM~\cite{boyd2011admm}, the optimization is decomposed into the following subproblems:
\begin{eqnarray}
k^{t+1}
& = &
\arg\min_k
\mathcal{D}\bigl(I_f, I_m \circ \phi_k\bigr)
+
\frac{\rho}{2}
\|D(k)-Z^t+U^t\|_F^2,\\
Z^{t+1}
& = &
\arg\min_Z
\lambda \eta \|Z\|_{2,1}
+
\frac{\rho}{2}
\|D(k^{t+1})-Z+U^t\|_F^2,\\
U^{t+1}
& = &
U^t + D(k^{t+1}) - Z^{t+1},
\end{eqnarray}
where $\rho > 0$ is the penalty parameter and $U$ is the scaled dual variable. The $k$-update corresponds to a smooth registration problem that can be solved efficiently with gradient-based optimization. The $Z$-update is a proximal step associated with the TV penalty and enforces sparsity in the deformation gradients. The dual update then accumulates the residual of the constraint $D(k)=Z$ and promotes consistency between the primal variables. In this way, ADMM provides an efficient and numerically stable framework for TV-regularized registration.

\subsection{Automatic Hyperparameter Tuning}
The method involves three main hyperparameters: the regularization weight $\lambda$, the LCC kernel bandwidth $w$, and the control-point spacing $K$~\cite{vishnevskiy2017isotropic}. These parameters strongly influence registration performance and must be adapted to the target dataset, since they depend on factors such as image modality, spatial resolution, noise, and the expected magnitude and type of motion. In this sense, they act as controls for transferring the method to new registration tasks.

The regularization weight $\lambda$ controls the trade-off between image matching and deformation regularity. If $\lambda$ is too small, the deformation may become under-regularized and less anatomically plausible; if it is too large, the solution may become over-regularized and suppress relevant local motion. The kernel bandwidth $w$ determines the spatial support of the Gaussian weighting kernel $H_w$ used in the LCC similarity term. Smaller values make the metric more local but also more sensitive to noise, whereas larger values improve robustness at the cost of reduced locality. The control-point spacing $K$ determines the flexibility of the deformation model. Larger values produce coarser control grids with fewer degrees of freedom, improving robustness and reducing computation but limiting local motion capture; smaller values allow finer and more detailed deformations but increase the number of parameters. The parametric representation therefore also acts as an implicit regularizer.

In~\cite{vishnevskiy2017isotropic}, the sensitivity of the optimal parameters was found to be low across four intra-patient breathing-motion datasets spanning both CT and MR. However, all of these datasets involved the same type of motion. When the method is applied to a substantially different task, such as inter-patient brain registration, or to a new benchmark with different image characteristics, suitable parameter values are not known in advance and manual tuning becomes impractical.

To address this, our framework consists of two stages: (i) automatic hyperparameter selection using Bayesian optimization, and (ii) registration of new image pairs using the selected parameters. As discussed, hyperparameters $(\lambda, w, K)$ must be adapted to each dataset. Manual tuning is time-consuming and risks overfitting, so we automate this step.
Specifically, Bayesian optimization with a Gaussian process surrogate and expected improvement acquisition function~\cite{snoek2012practical} searches the hyperparameter space. Each trial registers all available annotated pairs and returns a task-dependent evaluation metric, such as target registration error for landmark-based evaluation or Dice score for segmentation-based evaluation. After $N$ trials, the configuration with the best performance is selected and applied to the test set, which is not used during parameter selection.
To obtain an unbiased estimate of the tuning procedure itself, a nested cross-validation can additionally be performed: the annotated pairs are divided into $F$ folds and the Bayesian search is run independently within each fold using only the tuning pairs, with the held-out pairs serving for evaluation. The mean performance across folds then provides a less biased estimate of how well the automatic selection generalizes. When manual tuning is desired, the selected parameters also provide a sensible starting point.

\section{Experiments}
\subsection{Compared Methods}
Since our work is centered around pTVreg, we first distinguish the different variants considered in this study. The pTVreg method~\cite{vishnevskiy2017isotropic}, together with its released MATLAB implementation, is referred to here as \emph{pTVreg (original)}. We additionally include \texttt{imregdeform} from the MATLAB Medical Imaging Toolbox as a software baseline for total-variation-based deformable registration that is documented with reference to the method of Vishnevskiy et al.~\cite{mathworks_imregdeform,vishnevskiy2017isotropic}. Our own Python implementation is reported as \emph{pTVreg (ours)}, which uses the described ADMM-based optimization scheme. To assess the impact of this design choice, we also include an ablation variant, denoted \emph{pTVreg (ours\textbackslash ADMM)}, in which the same framework is optimized with L-BFGS using a smooth TV approximation instead of ADMM.

In addition to these pTVreg-related variants, we compare against classical registration methods, namely corrField and deeds, and against learning-based methods, namely VoxelMorph, cLapIRN, VoxelMorph+, VoxelMorph++, and SITReg~\cite{heinrich2015corrfield,heinrich2013deeds,balakrishnan2019voxelmorph,mok2020lapirn,heinrich2022voxelmorphpp,honkamaa2024sitreg}. Results for corrField, deeds, VoxelMorph, VoxelMorph+, and VoxelMorph++ are taken from the Lung250M-4B benchmark ~\cite{falta2023lung250m}, while SITReg and cLapIRN results are taken from Honkamaa et al.~\cite{honkamaa2024sitreg}.

\subsection{Dataset, Evaluation, and Implementation}
We compare the above methods on the Lung250M-4B benchmark for intra-patient inhale-exhale lung CT registration~\cite{falta2023lung250m}. 
This benchmark involves multiple datasets, one of which is COPDgene~\cite{castillo2013copdgene} containing 10 intra-patient inhale-exhale breathing CT volume pairs used for testing; as well as several other datasets (involving more than 100 additional volume pairs) which are provided for deep learning methods for training but are not used herein.
We evaluate registration accuracy using target registration error (TRE), computed as the mean Euclidean distance after registration between provided manual ground-truth landmark pairs in the COPDgene volumes.

For all pTVreg variants, including imregdeform, all images are clipped to the Hounsfield-unit range $[-1000,500]$, normalized to $[0,1]$.
Our implementation is based on PyTorch using GPU-accelerated computation of only the image metric. 
For the Lung250M-4B benchmark, we use the proposed Bayesian optimization to identify the parameter set
$\lambda{=}0.12$, $K{=}8$, and $w{=}2.1$ used in our results.

\subsection{Results}
\label{sec:quantitative}

\begin{figure}
\centering
\includegraphics[height=.4\textwidth]{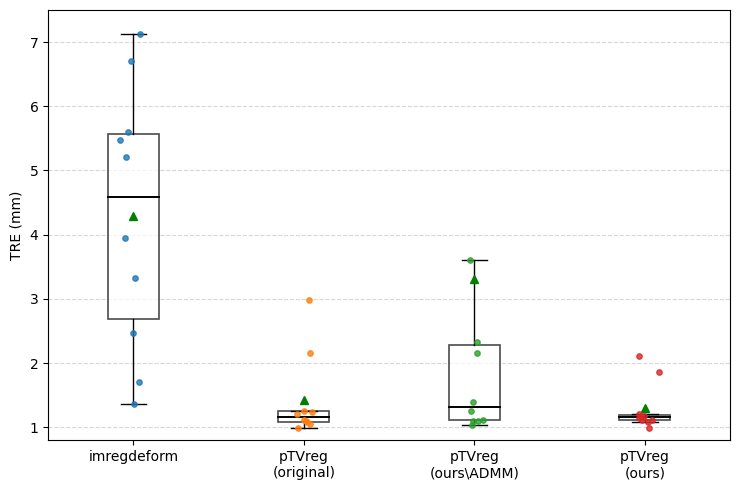}
\caption{Per-case TRE distribution for three pTVreg variants on Lung250M-4B. Each box summarizes the distribution across test cases.  
(For readability, the box plot is displayed in a zoomed range; the extreme outlier of pTVreg (ours\textbackslash ADMM) for case\_113, with TRE = 17.99 mm, is not shown in the figure.)}
\label{fig:boxplot_tre}
\end{figure}

Table~\ref{tab:lung_results} reports the test TRE on the Lung250M-4B
benchmark. 
\begin{table}
\centering
\caption{Comparison on the Lung250M-4B test cases using target registration error (TRE, mm). Standard deviations are shown in parentheses when available; ``(-)'' indicates that no standard deviation was reported in the original work.}
\label{tab:lung_results}
\begin{tabular}{ll@{\quad}l}
\hline
& Method & TRE \\
\hline
\multirow{5}{1em}{\rotatebox{90}{Learned}} & VoxelMorph$^{\dagger}$~\cite{balakrishnan2019voxelmorph}      & 6.66 (-) \\
& cLapIRN$^{\dagger}$~\cite{mok2020lapirn}         & 5.34 (1.90) \\
& VoxelMorph+ (w/ IO)~\cite{falta2023lung250m}         & 4.31 (-) \\
& VoxelMorph++ (w/ IO)~\cite{falta2023lung250m}       & 2.26 (-) \\
& SITReg$^{\dagger}$~\cite{honkamaa2024sitreg}                     & 2.71 (0.93) \\
\hline \hline
\multirow{6}{1em}{\rotatebox{90}{Analytical}} 
& corrField~\cite{heinrich2015corrfield}                   & 1.45 (-) \\
& deeds~\cite{heinrich2013deeds}                       & 1.53 (-) \\
& imregdeform~\cite{mathworks_imregdeform}           & 4.29 (2.04) \\
& pTVreg (original)~\cite{vishnevskiy2017isotropic}           & 1.41 (0.64) \\
\cline{2-3}
& pTVreg (ours\textbackslash ADMM)          & 3.30 (5.20) \\
& pTVreg (ours)               & 1.29 (0.37) \\
\hline
\end{tabular}
\vspace{0.5em}
\parbox{0.95\textwidth}{\footnotesize
$^{\dagger}$ The method is taken from the SITReg paper~\cite{honkamaa2024sitreg} reported in inhale coordinates, so it is not directly comparable to exhale-coordinate results.
}
\end{table}
Figure~\ref{fig:boxplot_tre} complements Table~\ref{tab:lung_results} by showing the per-case TRE distribution for the three pTVreg variants. 

As can be seen, \emph{pTVreg (ours)} achieves the lowest TRE values and the most compact overall distribution. In contrast, pTVreg (ours\textbackslash ADMM) shows a wider spread across cases and worse TRE. 
Possible reasons include differences in both the optimization procedure and the interpolation used for warping. The L-BFGS implementation in PyTorch may not behave identically to the \emph{minFunc}~\cite{schmidt2005minfunc} optimizer used in pTVreg (original), which may result in different optimization trajectories. In addition, pTVreg (ours) and pTVreg (ours\textbackslash ADMM) use PyTorch \texttt{grid\_sample} for image warping, which provides trilinear interpolation for 3D volumes, whereas pTVreg (original) uses tricubic interpolation. Because tricubic interpolation is not supported for 5-D inputs in PyTorch, this implementation difference may also affect performance; in our experiments, tricubic interpolation showed a beneficial effect on registration accuracy.

\begin{figure}
\centering
\includegraphics[width=\textwidth]{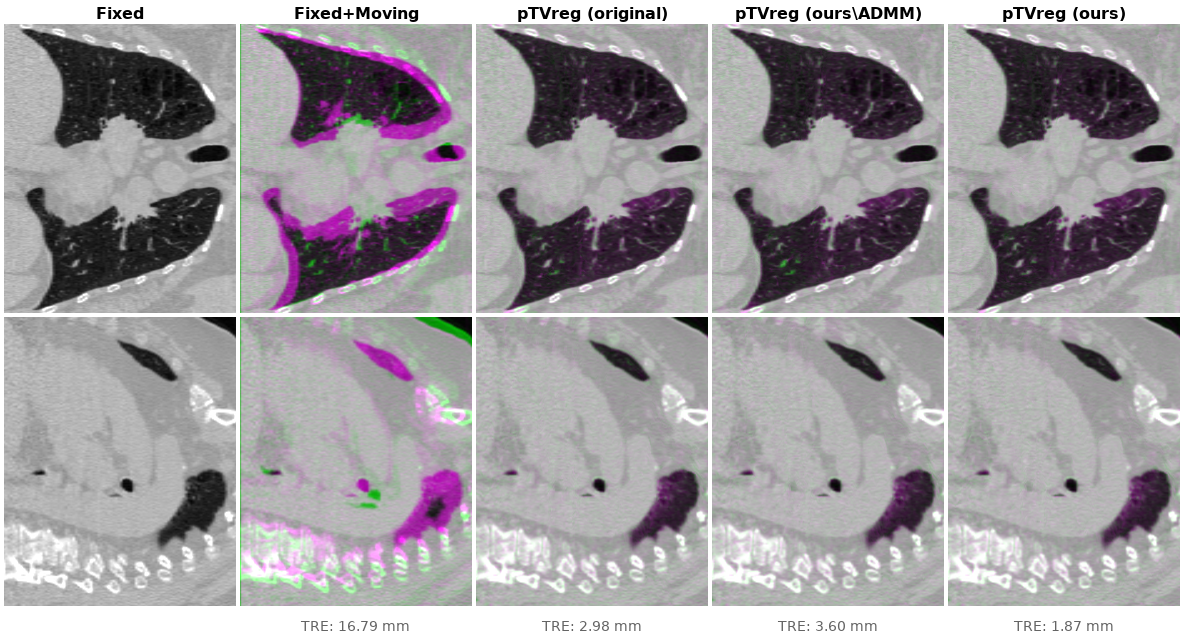}
\caption{Example registration for \emph{case\,105} from Lung250M-4B in coronal (top) and axial (bottom) views. The fixed image is shown in grayscale, while overlays visualize the fixed and moving images before and after registration. All pTVreg variants reduce the initial misalignment, and the proposed implementation achieves the lowest TRE for this case.}
\label{fig:case105}
\end{figure}

Figure~\ref{fig:case105} shows a representative example from Lung250M-4B (case 105). 
Compared with the initial fixed--moving overlay, all three pTVreg variants visibly reduce the misalignment in both coronal and axial views where \emph{pTVreg (ours)} yields the best quantitative result for this case and reduces the TRE from 16.79 mm before registration to 1.87 mm. The pTVreg (original) and pTVreg (ours\textbackslash ADMM) also significantly improve alignment, with final TRE values of 2.98 and 3.60 mm, respectively. These results are also consistent with the quantitative findings.

\subsection{Cross-Task Transfer}
To illustrate the importance of dataset-specific tuning, we
applied our method to inter-patient brain MRI registration
on the OASIS dataset from Learn2Reg~\cite{hering2023learn2reg,marcus2007oasis}, using the average Dice score over 35 anatomical labels over 19 validation cases as evaluation metric which measures segmentation overlap.
When using the Lung250M-4B parameters directly
($\lambda{=}0.12$, $w{=}2.1$, $K{=}8$), the method
achieves a mean Dice of 0.8025. After using our proposed automatic
hyperparameter selection process leading to the hyperparameter set \{$\lambda{=}0.10$, $w{=}4$, $K{=}2$\}, then the Dice score improves substantially to 0.8342.
This shows that settings found for one task are not the best choice for another task, and shows the practical advantage of choosing them automatically.
\section{Conclusion}
\label{sec:conclusion}
We introduce herein a Python reimplementation of pTVreg for
deformable image registration, including ADMM-based optimization and automatic
hyperparameter selection using Bayesian optimization.
In the Lung250M-4B benchmark, the proposed implementation achieved the best
performance among the compared methods, outperforming learning-based
baselines. These results show that an analytical method can still yield competitive and superior results to deep learning in a common deformable registration task.

\begin{credits}
\subsubsection{\ackname}
This work was supported by the Swedish Foundation for Strategic Research (SSF) under grant ID23-0018. The computations were enabled by resources provided by the National Academic Infrastructure for Supercomputing in Sweden (NAISS), partially funded by the Swedish Research Council through grant agreement no. 2022-06725.

\end{credits}
%
% ---- Bibliography ----
%
\bibliographystyle{splncs04}
\bibliography{arxiv}
\end{document}